\pdfoutput=1
\documentclass[journal,onecolumn,draftclsnofoot]{IEEEtran}

\usepackage{graphicx}
\usepackage{url}
\begin{document}
\title{The Parallel Distributed Image Search Engine (ParaDISE)}
\author{\IEEEauthorblockN{Dimitrios Markonis,
Roger Schaer,
Alba G. Seco de Herrera and
Henning M\"uller}

\IEEEauthorblockA{
University of Applied Sciences Western Switzerland (HES--SO), Business Information Systems,\\
TechnoArk 3, 3960 Sierre, Switzerland\\
Email: dimitrios.markonis@hevs.ch}
}
\maketitle
\begin{abstract}
Image retrieval is a complex task that differs according to the context and the user requirements in any specific field, for example in a medical environment.
Search by text is often not possible or optimal and retrieval by the visual content does not always succeed in modelling high-level concepts that a user is looking for.
Modern image retrieval techniques consists of multiple steps and aim to retrieve information from large--scale datasets and not only based on global image appearance but local features and if possible in a connection between visual features and text or semantics.

This paper presents the Parallel Distributed Image Search Engine (ParaDISE),
an image retrieval system that combines visual search with text--based retrieval and that is available as open source and free of charge.
The main design concepts of ParaDISE are flexibility, expandability, scalability and interoperability.
These concepts constitute the system, able to be used both in real--world applications and as an image retrieval research platform.

Apart from the architecture and the implementation of the system, two use cases are described, an application of ParaDISE in retrieval of images from the medical literature and a visual feature evaluation for medical image retrieval.
Future steps include the creation of an open source community that will contribute and expand this platform based on the existing parts.
\end{abstract}
\section{Introduction}
Image retrieval just like general information retrieval is a popular and frequent activity in many fields such as journalism~\cite{MaS1998} and medicine~\cite{MHD2012}.
In certain cases, describing with keywords the images to retrieve is often not possible or optimal.
Content--based image retrieval (CBIR) is an alternative approach to image search that uses the visual content of the image to find similar images.
Querying by image example can be really time efficient, especially with the use of user interaction techniques such as relevance feedback~\cite{Roc1971}, which allows quick query refinement by marking relevant results.
However, due to the use of low--level visual characteristics, such as color, shape and texture, by CBIR in order to represent an image, it is difficult to describe high--level concepts, e.g. a pathology found in an X--ray.
This is particularly important in difficult cases, e.g. medical image retrieval where abnormalities and pathologies may be found in small areas of the image.
Multi--modal approaches are one way to cope with this ``semantic gap'', combining text and visual information to determine relevancy to the query~\cite{AHE2010}.

Research on CBIR has been carried out in several fields such as object and scene retrieval~\cite{SiZ2003} and remote sensing~\cite{DDP2003}.
In the early years, mathematical models where used to represent the visual content of the image in a holistic manner~\cite{OlT2001}.
Later, local descriptors~\cite{Low2004} modelling the information around specific points or ROIs were shown to outperform global descriptors in several tasks~\cite{MiS2005,DKN2004}.
While local descriptors allowed for partial matching of images and showed scale and rotation invariance, they were inefficient for search within large--scale image collections.
For this reason, more compact representations inspired from text--based information retrieval such as Bag--of--Visual-Words (BoVW)~\cite{SiZ2003} have been developed.
Efficient indexing structures such as the Inverted Index have also been employed to allow for fast real--time search~\cite{MSM1999c}.

Several projects have already been realized in the field of information retrieval and made systems available as open source.
Among them is the Viper project~\cite{SMM2001}, the outcome of which was the GNU Image--Finding Tool (GIFT), a CBIR system that enables users to perform ``Query By Example'' search operations and improve the quality of results using relevance feedback. 
The system contained a relatively small bank of outdated visual features which was hard to modify and expand.

Another noteworthy project is Lucene Image Retrieval (LIRe)~\cite{LuC2008}, a library based on the Lucene text retrieval software, which provides various visual features.
The system uses purely visual search and provides little support for several state--of--the--art representations (such as spatial pyramid matching or bag--of--colors), indexing parallelization or flexible index structuring.

Flexible Image Retrieval Engine (FIRE) is another example of a CBIR system~\cite{DKN2008}, also used in medical image retrieval evaluation apart from other applications.
The system allows also combination with text queries.
Being developed before 2007, the system does not support state--of--the--art mid--level representations (such as BoVW or Vectors of locally aggregated descriptors (VLAD)~\cite{JDS2010}).
No parallelization schema is mentioned for indexing large scale datasets, either.

In~\cite{YKA2009} a CBIR system, NIR, Nutch~\cite{KCS2004} and LIRe is presented.
It uses Hadoop~\cite{Whi2010}, which is an implementation of the MapReduce framework~\cite{DeG2008}, for parallel computing.
A small bank of outdated features is used to demonstrate the system using Hadoop.
MapReduce was also used for the online processes even though this is not advised~\cite{LLC2011}.
The indexing and retrieval times were demonstrated in a relatively small database.

Another system called Distributed Image Retrieval System (DIRS) is described in~\cite{ZLL2010} using LIRe and HBase(\footnote{\texttt{http://hbase.apache.org/}}).
Data sets of up to 100,000 images are used for testing the query times.
When using datasets above 20,000 images, the retrieval times reported are restrictive for online use even though they are faster than without Hadoop use.

This study presents the Parallel Distributed Image Search Engine (ParaDISE). ParaDISE is an image retrieval system that combines CBIR and text--based retrieval.
The design of the system was based on the difficult use case of medical image retrieval, after a survey on radiologists image search information needs~\cite{MHD2012}.
The design concepts are, however, relevant to any image retrieval field.
ParaDISE constitutes a platform that could be used both in research, for CBIR and multi--modal image retrieval, but also in large--scale applications.
The design and implementation of ParaDISE is described in Section~\ref{sec:system}.
Two use cases demonstrating the applications of ParaDISE are presented in Section~\ref{sec:usecases}.
The system design concepts and implementation choices are discussed in Section~\ref{sec:discussion}.

\section{System Description}\label{sec:system}
In this section, the findings of the survey carried out on visual information search~\cite{MHD2012} are discussed and translated into a list of system requirements.
Then, the design and the implementation of a novel image retrieval system, named Parallel Distributed Image Search Engine (ParaDISE) are described in detail.

\subsection{Specifications and System Design}\label{sec:specs}

The observations of the workflow in the investigation of the image search behavior showed that the need for additional information during clinical duties occurred when the pathology of an abnormality found in a new case was unclear or unknown.
Moreover, it was often mentioned in the survey that images or interesting cases were searched for lectures or presentations in academic work.
Thus, the radiologist may or may not know some keywords to initiate the search.
This dictates that a medical image retrieval system should support querying by text, by image example (e.g for the cases where no pathology keywords are known) or a combination of the two (e.g for cases that the user may have a hint but not certainty).
Relevance feedback or term suggestion techniques could also help refine the search if the object of the search is not fully clear.

The Internet was mentioned as one of the main sources where radiologists where seeking for information.
At the same time, the quality of the results and the case context associated with the image were mentioned among the most important criteria when judging the results' quality.
As peer--reviewed articles can be considered as a trustworthy source, indexing images from the medical literature on the Internet can achieve a high level of result quality and quantity.
A search system should provide linking and easy navigation between the images and their associated case.

Linking of internal sources, such as PACS with the medical literature and personal archives was also considered important when searching for information.
As these sources contain heterogeneous imaging data, different features and image representations need to be supported.
Extending the search into multiple indices should be possible, as the ability to interconnect with other search systems.

The main reason for image search failure given by the participants of the survey was that the information sought was too rare.
However it was believed that it should have been available somewhere but they couldn't find it.
Moreover search needs to be fast as radiologists have very tight schedules.
In order to provide quick access to new findings on a rapidly--growing scientific field, the system needs to have regular index updates and be scalable to millions of images and articles.

A first list of system requirements can be derived from this analysis:

\begin{itemize}
 \item support of query by keywords, image example or combination of both;
 \item index of a trustworthy source, such as the medical literature on the Internet;
 \item linking of images and associated articles, easy navigation between the two;
 \item support of different visual features and image representations;
 \item support of search into multiple indices, interoperability;
 \item scalability and support of regular index updates.
\end{itemize}

\subsection{Architecture}\label{sec:paradise}

The design of ParaDISE architecture was based on the following concepts: flexibility, expandability and scalability.
The development was split into two parts, the Backend and the Frontend, and are described in the following:

\paragraph{\textbf{Backend}}
The ParaDISE backend follows an object--oriented architecture and consists of basic elements, called Components.
Each Component is associated with a Manager object.
The Manager is responsible for selecting one out of the supported instances of the Component.
The behavior of the selected instance is controlled by a Parameters object that contains the tunable values of the method implemented in the instance.
The Components are:
\begin{itemize}
\item \textit{The Extractor:}\\ undertakes the extraction of local descriptors.
More information on the local feature extractors supported in the Extractor can be found in Section~\ref{sec:local_feats_2d}.
\item \textit{The Descriptor:}\\ is responsible for the mid--level features aggregating the local descriptors extracted by the Extractor.
It also contains global descriptors, for which no local features extraction is needed.
More information on the global descriptors and mid--level features supported in the Descriptor exists in Section~\ref{sec:global_feats_2d}.
\item \textit{The Storer:}\\ is used to store the image representation vectors produced by the Descriptor during the indexing process.
It is also responsible for accessing the index during online search.
The storing methods supported in the Storer are described in Section~\ref{sec:paradise_storer}.
\item \textit{The Fusor:}\\ undertakes the fusion of retrieved results lists.
These can be either lists retrieved by multiple image queries or results retrieved using different features, indices and even other image retrieval systems.
The fusion rules supported in the Fusor are described in Section~\ref{sec:paradise_fusor}.
\end{itemize}

 \begin{figure}[!tp]
   \begin{center}
     \includegraphics[width=13cm]{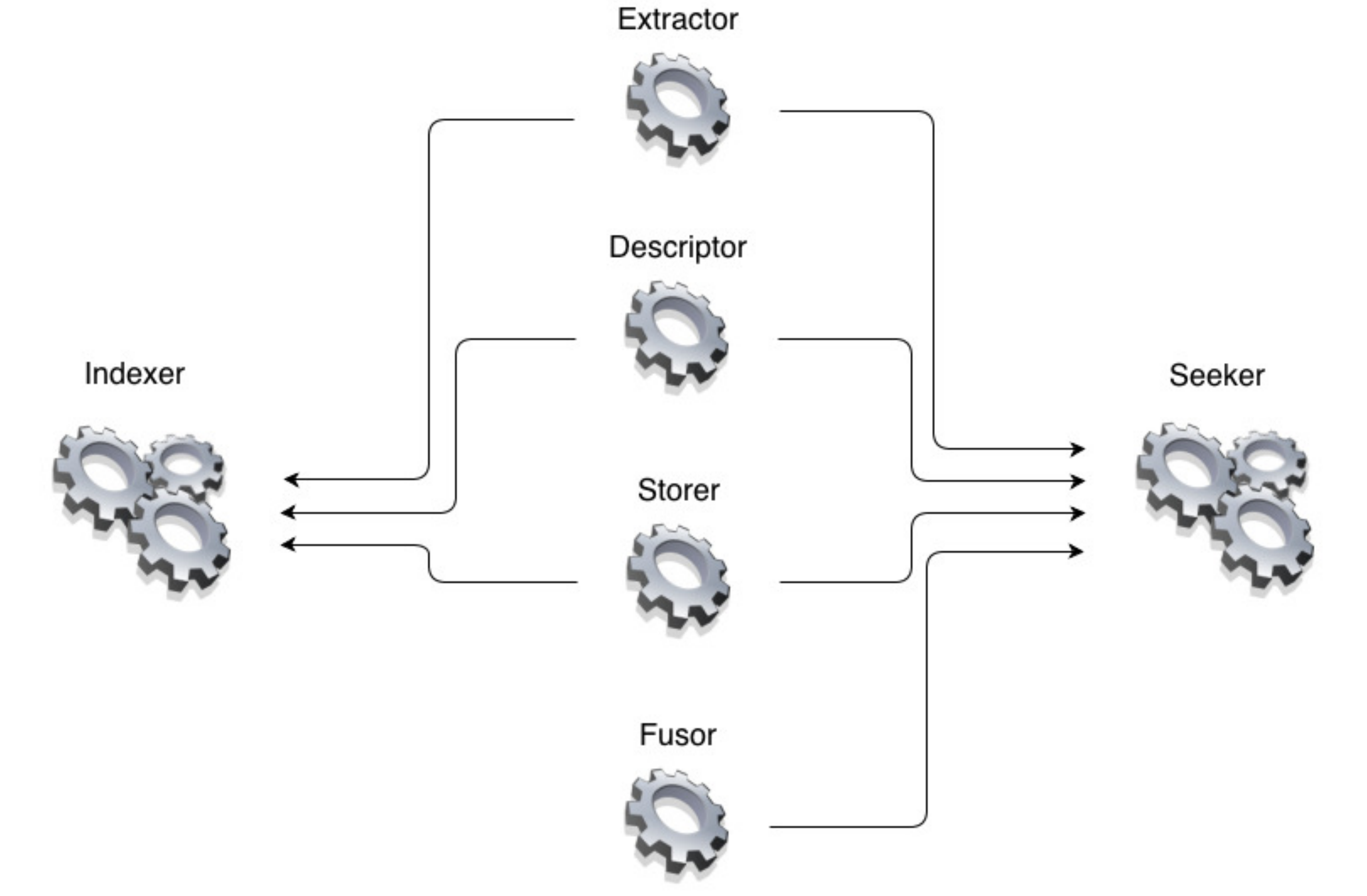}
 \caption{An overview of the ParaDISE backend. The four basic elements are combined to perform the indexing and search processes.\label{fig:ParaDISE_backend}}
  \end{center}
 \end{figure}

The Components are combined to perform the two main operations for CBIR, offline indexing of the database images and online search using a set of image examples (Figure~\ref{fig:ParaDISE_backend}).
These two processes are implemented in complex ParaDISE elements, called CompositeComponents: the Indexer and the Seeker.
Again, a Manager is used to select an available Indexer or Seeker and a CompositeParameters object is used to control its behavior.
The indexing and search processes are described in more detail in Sections~\ref{sec:paradise_index} and~\ref{sec:paradise_retrieval} respectively.

\begin{figure}[!tp]
   \begin{center}
     \includegraphics[width=9cm]{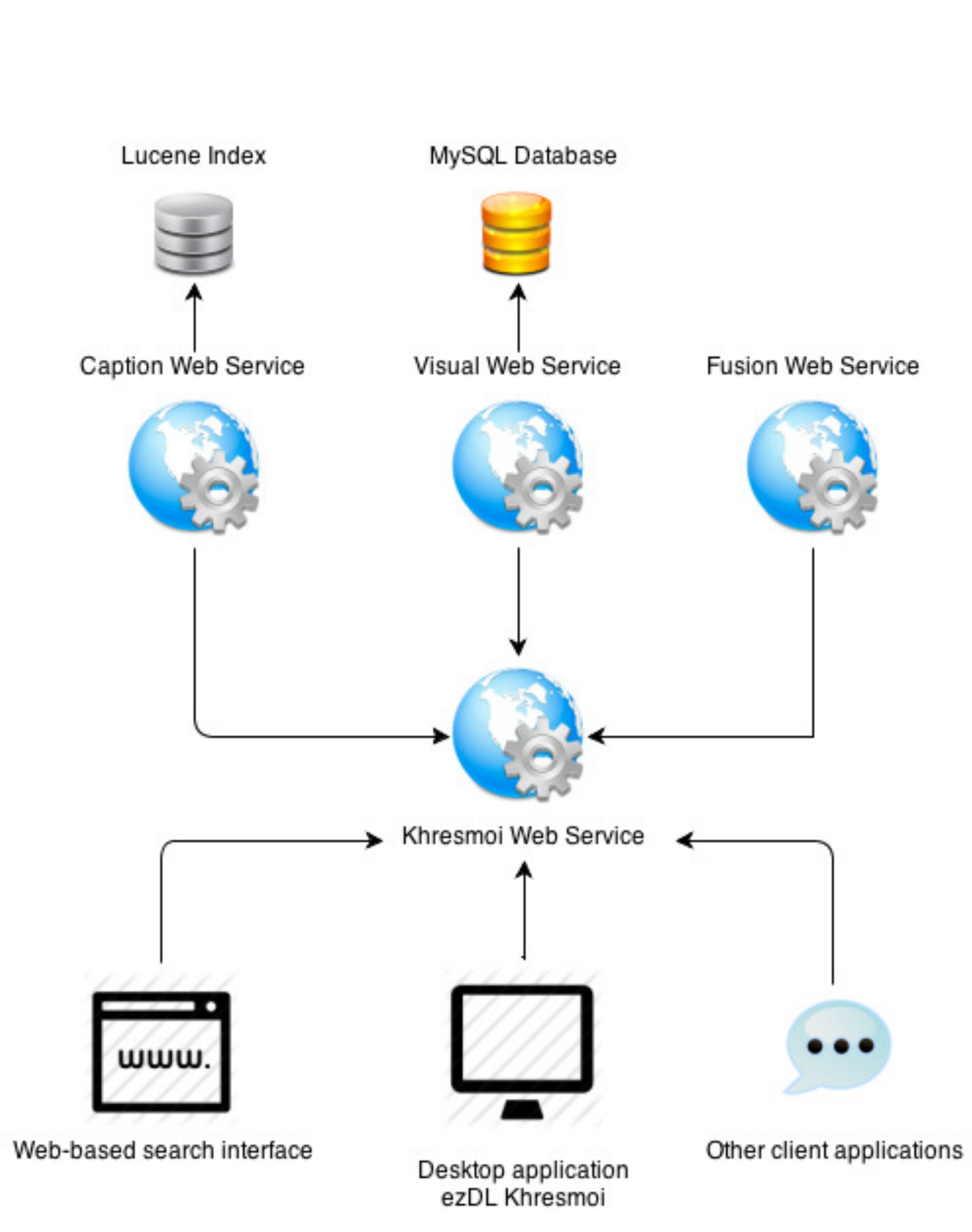}
 \caption{An overview of the ParaDISE frontend. Web services for visual and textual search are combined using the Fusion Web service. The global Web service serves as an interface point to external client applications.\label{fig:frontend}}
  \end{center}
 \end{figure}

The Components count was kept as low as possible to cover most CBIR approach pipelines without making the system architecture too complex.
However, the addition of new Components (e.g. a Preprocessor or a Classifier) is relatively simple due to the component--based architecture.
JAVA was chosen as the main programming language for the implementation of the ParaDISE backend.
\paragraph{\textbf{Frontend}}
The ParaDISE frontend, namely the service layer, consists of multiple Web services which use a REpresentational State Transfer (REST)--style  architecture (Figure~\ref{fig:frontend}).
Standard Hyper Text Transfer Protocol (HTTP) GET and POST requests are used to communicate with the Web services.
However, an offline version of ParaDISE frontend exists in the form of a JAVA library.
This facilitates easy installation and usage of the engine for single--server applications, personal databases and small--scale research experiments.

A large bank of visual feature extractors has been built into the ParaDISE system.
These features are split into two categories, local features and global descriptors, and are presented in Sections~\ref{sec:local_feats_2d} and~\ref{sec:global_feats_2d}

\subsubsection{The Extractor}\label{sec:local_feats_2d}
Local features have been used in CBIR for more than a decade~\cite{Low2004}, demonstrating state--of--the--art performance in many applications~\cite{MiS2005,CaJ2005}.
They represent low--level visual characteristics of regions of the image, such as color, shape and texture.
The local feature extraction takes place in the Extractor component of ParaDISE.
The following local features are supported in the current version of ParaDISE (see also~\cite{LBO2012}):
\begin{itemize}
\item \textit{Scale Invariant Feature Transform (SIFT)}~\cite{Low2004}

(The implementation of the SIFT feature in the Fiji image processing package\footnote{\url{http://fiji.sc/}{http://fiji.sc/}} was used.)

\item \textit{Speeded Up Robust Feature (SURF)}~\cite{BTG2006}
(The implementation of the SURF feature in the Fiji image processing package was used.)
\item \textit{RootSIFT}~\cite{ArZ2012}
\item \textit{Lab local features}~\cite{WDJ2011}

\end{itemize}

\subsubsection{The Descriptor}\label{sec:global_feats_2d}

While local features perform well in object recognition, image classification and CBIR, they are inefficient for large--scale tasks.
For this reason statistical image representations have been used, also called mid--level features, with BoVW~\cite{SiZ2003} being the most commonly used.
Moreover, since there is no one--solution--fits--all in image retrieval applications, other global descriptors have been included in the feature bank.
The following mid--level features and global descriptors are supported in the Descriptor component of ParaDISE:

\begin{itemize}
\item \textit{BoVW}~\cite{SiZ2003};

The following variants of BoVW are available:
\begin{itemize}
\item \textit{Binary BoVW~}\cite{SiZ2003};
\item \textit{Grid BoVW};

\item \textit{Spatial Pyramid Matching (SPM) BoVW}~\cite{LSP2006};
\end{itemize}
\item \textit{Vector of Locally Aggregated Descriptors (VLAD)}~\cite{JDS2010};
\item \textit{GIST}~\cite{OlT2001}
(the implementation provided in~\cite{OlT2001} was used);
\item \textit{Riesz miniature}~\cite{DFV2012a}
(an adapted version of the implementation provided in~\cite{DFV2012a} was used);
\item \textit{Histograms of Gradients (HoG) miniature}~\cite{DHB2011};
\item \textit{Gabor Filters}~\cite{SMM1999};
\item \textit{Tamura}~\cite{TMY1978}
(for the implementation of this feature, the LIRe library was used~\cite{LuC2008});
\item \textit{Color and Edge Directivity Descriptor (CEDD)}~\cite{ChB2008a}
(the implementation in LIRe was used);
\item \textit{Fuzzy Color and Texture histogram (FCTH)}~\cite{ChB2008b}
(the implementation in LIRe was used);
\item \textit{Color Layout}~\cite{KaY2001}
(the implementation in LIRe was used);
\item \textit{Fuzzy Color histogram}~\cite{HaM2002}
%
%
(the implementation in LIRe was used);%
\item \textit{HSV Color histogram};
(the implementation in LIRe was used);
\item \textit{Singular Value Decomposition (SVD)}~\cite{OAD2011}.
\end{itemize}

\subsubsection{ParaDISE Storer}\label{sec:paradise_storer}

Four different Storers are currently supported in ParaDISE:
\begin{itemize}
 \item \textit{CSV Storer}\\
This Storer uses a Comma--Separated Values (CSV) file to store the index.
It is mostly suitable for research evaluations and small image collections, as it is very inefficient for real applications.
 \item \textit{SQL Storer}\\
The SQL storer stores the image descriptor vectors in a table in a MySQL database.
It can be used for application use cases and can handle large datasets as well as image vectors of small dimensionality.
 \item \textit{CouchDB Storer}\\
A noSQL alternative of SQL storer for image vectors of high dimensionality, such as concatenated feature vectors or BoVW models with large vocabularies.
 \item \textit{Cassandra Storer}\\
Cassandra Storer stores the index in a column family of a Cassandra\footnote{\url{http://cassandra.apache.org/}{http://cassandra.apache.org/}} keyspace.
Cassandra allows to have a parallel database with millions of columns.
This makes it suitable for very large datasets and image vectors of very high dimensionality.
\end{itemize}
\subsubsection{The Fusor}\label{sec:paradise_fusor}
The fusion rules supported in Fusor are:
 \begin{itemize}
  \item  \textit{CombSUM}
\begin{equation}
S_{\texttt{\footnotesize combSUM}}(i)=\sum_{k=1}^{N_k}{S_k(i)}
\end{equation}
where $S_k(i)$ is the score assigned to image $i$ in retrieved list $k$.
  \item  \textit{CombMNZ}
\begin{equation}
 S_{\texttt{\footnotesize combMNZ}}(i)=F(i)* S_{\texttt{\footnotesize combSUM}}(i)
\end{equation}
where $F(i)$ is the number of times an image $i$ is present in retrieved lists with a non--zero score.
 \item  \textit{CombMAX}
\begin{equation}
S_{\texttt{\footnotesize combMax}}(i)= \max_k{S_k(i)}
\end{equation}
where $S_k(i)$ is the score assigned to image $i$ in retrieved list $k$.
 \item  \textit{CombMIN}
\begin{equation}
S_{\texttt{\footnotesize combMin}}(i)= \min_k{S_k(i)}
\end{equation}
where $S_k(i)$ is the score assigned to image $i$ in retrieved list $k$.
  \item  \textit{Linear Weighting}
\begin{equation}
S_{\texttt{\footnotesize linear}}(i)=\sum_{k=1}^{N_k}w_k{S_k(i)}
\end{equation}
with $w_k\in[0,1]$ and $ \sum_{k=1}^{N_k}w_k = 1$.
  \item  \textit{Borda Count}
\begin{equation}
S{\texttt{\footnotesize Borda}}(i) = \sum_{k=1}^{N_k} \frac{1}{R_k(i)}
\end{equation}
where $R_k(i)$ the rank of the image in retrieved list $k$.
 \item  \textit{Reciprocal Rank}
\begin{equation}
S{\texttt{\footnotesize RRF}}(i) = \sum_{k=1}^{N_k} \frac{1}{c+R_k(i)}
\end{equation}
where $c$ a constant and $R_k(i)$ the rank of the image in retrieved list $k$.
\end{itemize}

\subsubsection{The Indexer}\label{sec:paradise_index}
The indexing of the visual content of the image collection is an offline operation.
As mentioned in Section~\ref{sec:paradise}, the Indexer CompositeComponent is responsible for this task in ParaDISE.
Apart from serial indexing, parallel indexing is also supported using the MapReduce framework.
Below follows the description of the two currently supported methods:
 \begin{figure}[!tp]
   \begin{center}
     \includegraphics[width=12cm]{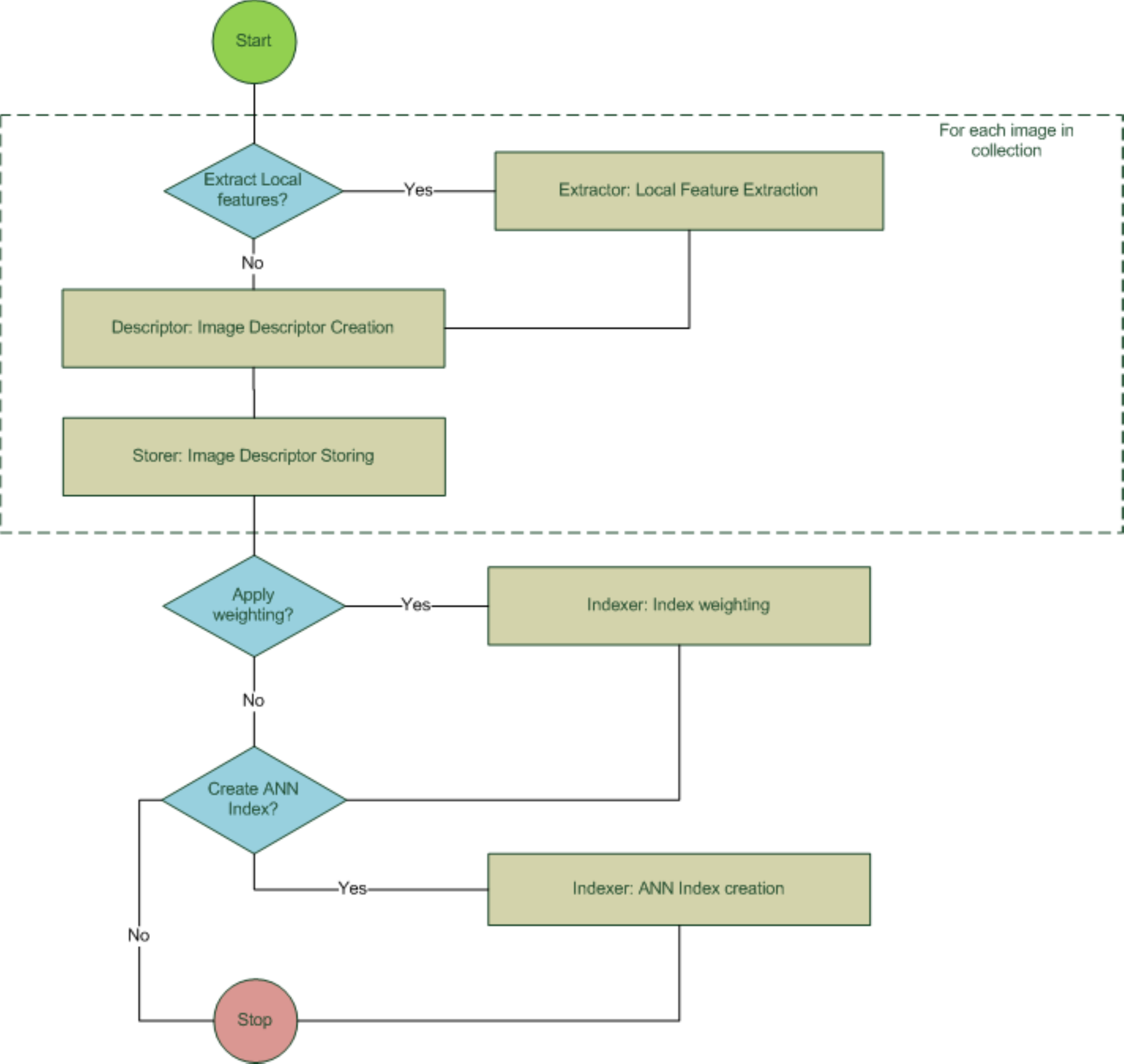}
 \caption{The indexing pipeline of ParaDISE Indexer.\label{fig:indexing_pipeline}}
  \end{center}
 \end{figure}
\begin{itemize}
 \item \textit{Serial Indexer}\\
The serial indexing pipeline uses the basic ParaDISE Components (see Figure~\ref{fig:indexing_pipeline}).
First, the local features of each image are extracted by the Extractor, if needed.
Then, the image descriptor is created by the Descriptor, either integrating the local feature vectors into a mid--level representation or using a global descriptor.
The Storer inserts that image descriptor vector into the index.
The direction in the decision nodes is decided by the values of the Indexer Parameters.

After the index is created, a weighting can be applied to the index.
The following weighting methods are supported:
\begin{itemize}
\item \textit{Term Frequency -- Inverse Document Frequency (TF--IDF)}\\
The TF--IDF weighting is widely used in text--based information retrieval.
The rationale behind this weighting is that words that are found often in a document contain more information.
At the same time, words that are found often in the document collection are not that informative.
The mathematical expression of TF--IDF  is the following:
\begin{equation}\label{equ:tfidf}
tfidf = \frac{n_{id}}{n_d}\log\frac{N}{n_i}
\end{equation}
where $n_{id}$ is the number of occurrences of word $i$ in document $d$,
$n_d$ is the total number of words in the document $d$,
$n_i$ is the number of occurrences of word $i$ in the whole database
and $N$ is the number of documents in the whole database.\\
It can be used in CBIR in combination with BoVW approaches.
\item \textit{Frequent Item Selection~\cite{VCJ2012}}\\
This weighting uses only the top $k$ TF--IDF values per image, called frequent items, to provide a compact image representation.
The images are then ranked according to the number of common shared frequent items with the query image.
\end{itemize}
Finally, the indexer can create an Approximate Nearest Neighbour (ANN) index structure to facilitate fast retrieval.
Currently, serial and parallel versions of Euclidean Locally Sensitive Hashing (E2LSH)~\cite{AnI2006} ANN method are supported.
This algorithm uses families of hashing functions to partition the index feature space and thus limit the search into the subspace that a query falls into.
\begin{figure}[!tp]
   \begin{center}
     \includegraphics[width=12 cm]{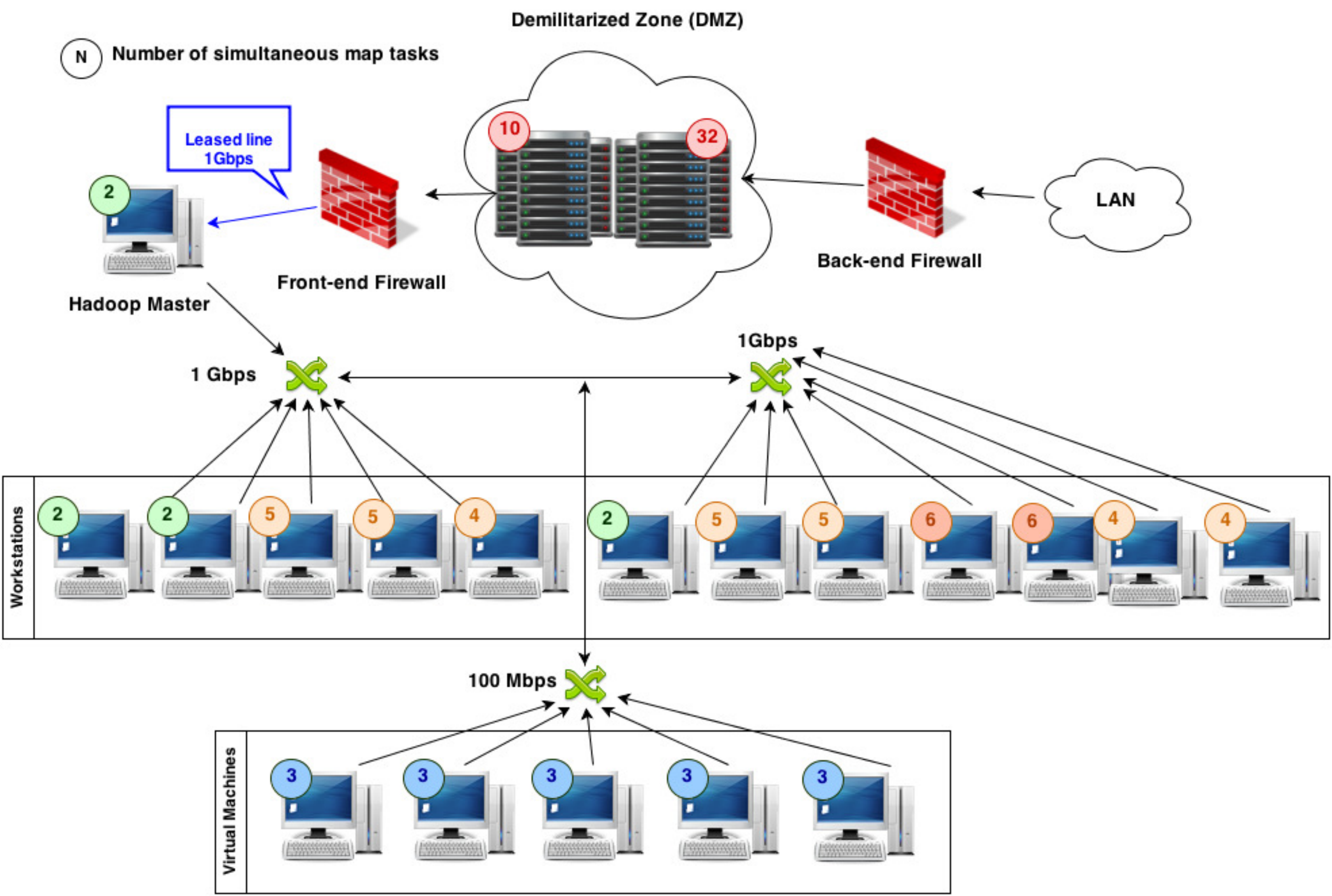}
 \caption{An overview of the HES--SO in--house Hadoop cluster.\label{fig:hadoop_cluster}}
  \end{center}
 \end{figure}
\item \textit{Hadoop Indexer}\\
The Hadoop~\cite{Whi2010} implementation of MapReduce was used for the parallelization of the indexing, since it is an easily parallelizable task.
The pipeline is identical to the one shown in Figure~\ref{fig:indexing_pipeline} except for the fact that the blocks in the frame are executed in parallel.
This is achieved by splitting the image collection into small groups of images.
Each group is indexed by a different map task.

Either an in--house or a cloud Hadoop cluster can be used for this indexing method, since the implementation is fully parametrizable.
An in--house cluster was created for the needs of the prototype, consisting of 13 workstations, 2 servers and 5 virtual machines.
This resulted in a 20 node cluster with a computational capability of 99 concurrent map tasks (Figure~\ref{fig:hadoop_cluster}).

The background of the framework and the details of the implementation of the cluster are described in~\cite{MSE2012a}.

\end{itemize}
Once the index is stored, the index parameters are saved in JSON format in a configuration file.
This way, the ParaDISE Seeker can use the same configuration for extracting the visual features of the query images when searching within the specific index.

\subsubsection{The Seeker}\label{sec:paradise_retrieval}
As mentioned in Section~\ref{sec:paradise}, the Seeker Composite Component is responsible for CBIR search in ParaDISE.
As required by CBIR, the ParaDISE Seeker allows similarity search using image examples as queries.
Multiple query images and negative examples are also supported.
From the user side, this allows for iterating the search using relevance feedback~\cite{Roc1971}.
The relevance feedback can be handled in various ways.
In ParaDISE Seeker the following algorithms are supported for handling relevance feedback:

 \begin{figure}[!tp]
   \begin{center}
     \includegraphics[width=12cm]{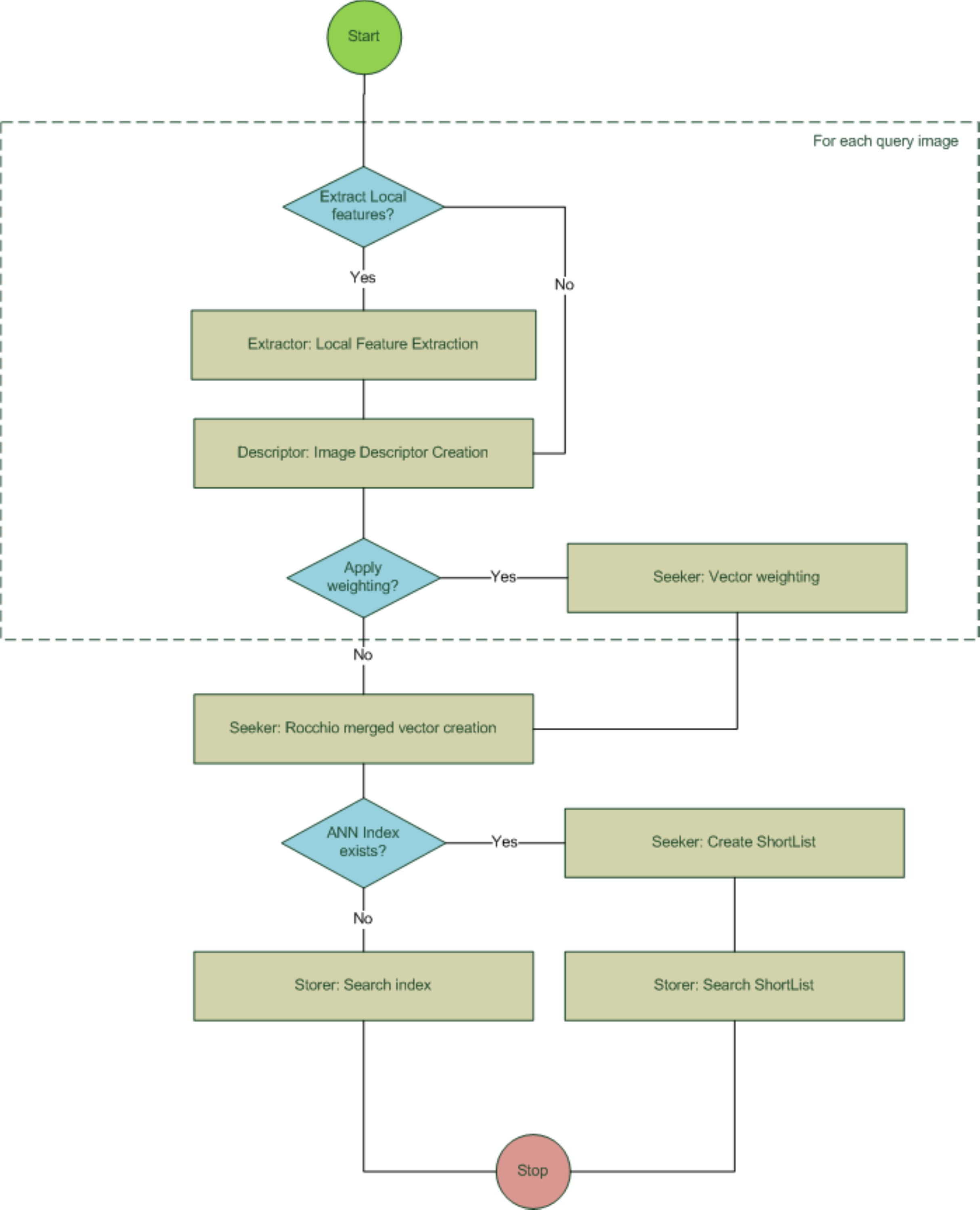}
 \caption{The search pipeline of the Rocchio Seeker.\label{fig:rocchio_pipeline}}
  \end{center}
 \end{figure}

\begin{itemize}
 \item \textit{Rocchio Seeker}\\
This Seeker uses the Rocchio algorithm~\cite{Roc1971} to handle multiple images of positive or negative relevance.
The Rocchio formula is given by:
\begin{equation}
 \vec{q}_m =
\alpha \vec{q}_o +
\beta \frac{1}{|D_r|} \sum_{\vec{d}_j\in D_r}\vec{d}_j -
\gamma \frac{1}{|D_{nr}|} \sum_{\vec{d}_j\in D_{nr}}\vec{d}_j
\end{equation}
where $\alpha,\beta$ and $\gamma$ are weights,
$\vec{q}_m$ is the modified query,
$\vec{q}_o$ is the original query,
$D_r$ is the set of relevant images and $D_{nr}$ is the set of non--relevant images.

The search pipeline of this method is shown in Figure~\ref{fig:rocchio_pipeline}.
The Seeker reads the index Parameters from the configuration file of the index it tries to access (see Section~\ref{sec:paradise_index}).
According to these Parameters, it transforms the images to the appropriate vector representations.
The Rocchio formula is then executed, producing a single merged vector.
If an ANN index exists for the accessed visual index then a shortlist of the vectors existing in the same subspace as the merged vector is returned.
In this case, the Storer searches within the returned shortlist, otherwise the whole index is searched.
The similarity search uses a distance metric or a similarity measure to rank the images.
The following distances/similarities are supported in ParaDISE:
 \begin{itemize}
  \item  \textit{Euclidean distance (L2 norm)}\\
\begin{equation}
      d_{\varepsilon}(\vec{p},\vec{q}) = \sqrt{\sum_{i=1}^n (p_i - q_i)^2}
\end{equation}
  \item  \textit{Manhattan distance (L1 norm)}\\
\begin{equation}
      d_{manhattan}(\vec{p},\vec{q}) = \sum_{i=1}^n |p_i - q_i|
\end{equation}
  \item  \textit{Canberra distance}\\
\begin{equation}
      d_{canberra}(\vec{p},\vec{q}) = \sum_{i=1}^n \frac{|p_i - q_i|}{|p_i| + |q_i|}
\end{equation}
  \item  \textit{$\chi^2$ distance}\\
\begin{equation}
      d_{\chi^2}(\vec{p},\vec{q}) = \frac{1}{2}\sum_{i=1}^n \frac{(p_i - q_i)^2}{p_i + q_i}
\end{equation}
  \item  \textit{Jeffrey divergence}\\
\begin{equation}
      d_{jd}(\vec{p},\vec{q}) = \sum_{i=1}^n (\log\frac{2p_i}{p_i + q_i} + \log\frac{2q_i}{p_i + q_i})
\end{equation}
\item  \textit{histogram intersection}\\
\begin{equation}
      s_{hi}(\vec{p},\vec{q}) = \sum_{i=1}^n \min(p_i,q_i)
\end{equation}
\item  \textit{Cosine similarity}\\
\begin{equation}
      s_{cosine}(\vec{p},\vec{q}) =  \frac{\sum_{i=1}^n(p_i \times q_i)}{||\vec{p}||\times||\vec{q_i}||}
\end{equation}
\end{itemize}
where $\vec{p},\vec{q}\in R^n$.

Also special similarity measures are supported for specific approaches:
 \begin{itemize}
  \item  \textit{Hamming Distance}\\
For binary vectors $\vec{p},\vec{q}$, the hamming distance $d(\vec{p},\vec{q})$ is defined as the number of ones of $p\oplus q$.
It can be used for comparing binary representations, such as binary BoVW.
  \item  \textit{Frequent Item Selection Distance}\\
This similarity is used in combination with the Frequent Selection weighting (see Section~\ref{sec:paradise_index}).
The similarity score is equal to the number of common shared frequent items.
\end{itemize}
 \item \textit{LateFusion Seeker}\\
 \begin{figure}[!tp]
   \begin{center}
     \includegraphics[width=12cm]{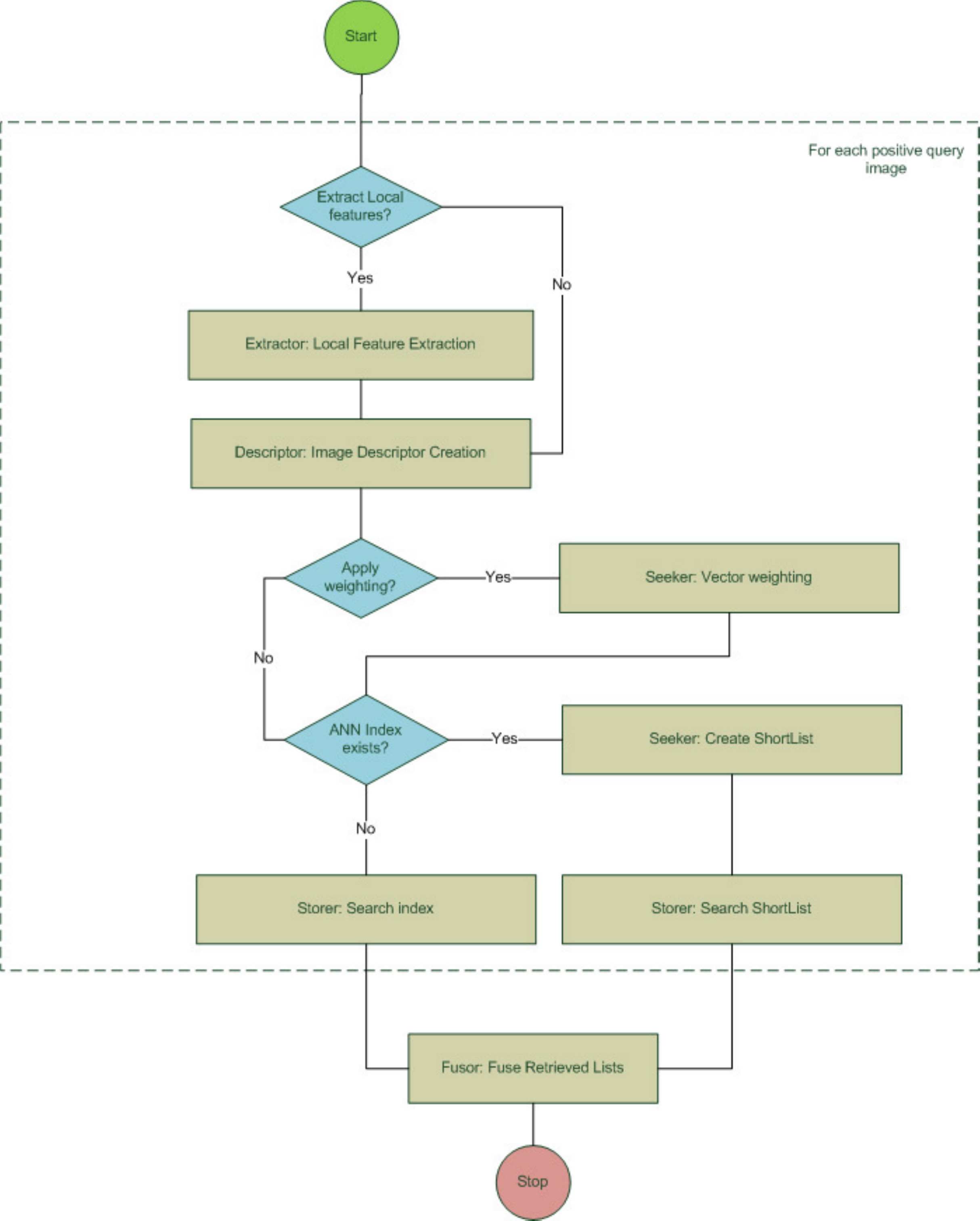}
 \caption{The search pipeline of the LateFusion Seeker.\label{fig:latefusion_pipeline}}
  \end{center}
 \end{figure}
The pipeline of this Seeker is demonstrated in Figure~\ref{fig:latefusion_pipeline}.
It is similar to the Rocchio Seeker pipeline but instead of producing a single merged query vector it initiates a different search for each positive query image.
In the end the Fusor Component is used to fuse the retrieved lists.
Negative query image examples are ignored.

\end{itemize}

\section{Use cases}\label{sec:usecases}
\subsection{The KHRESMOI medical literature access system}
KHRESMOI\footnote{\texttt{http://www.khresmoi.eu/}} is a project that aims at creating a multilingual, multi--modal search and access system.
One of the main functionalities of the system is to allow efficient access to the visual information available in electronic records and the open access medical literature on the Internet.
The system applies several novel information extraction and retrieval techniques, such as CBIR, relevance feedback and the use of the Semantic web in 2D and 3D medical image search.
ParaDISE is integrated in the KHRESMOI system, undertaking the task of searching for images and cases found in the open access medical literature.

 \begin{figure}[!tp]
   \begin{center}
     \includegraphics[width=13cm]{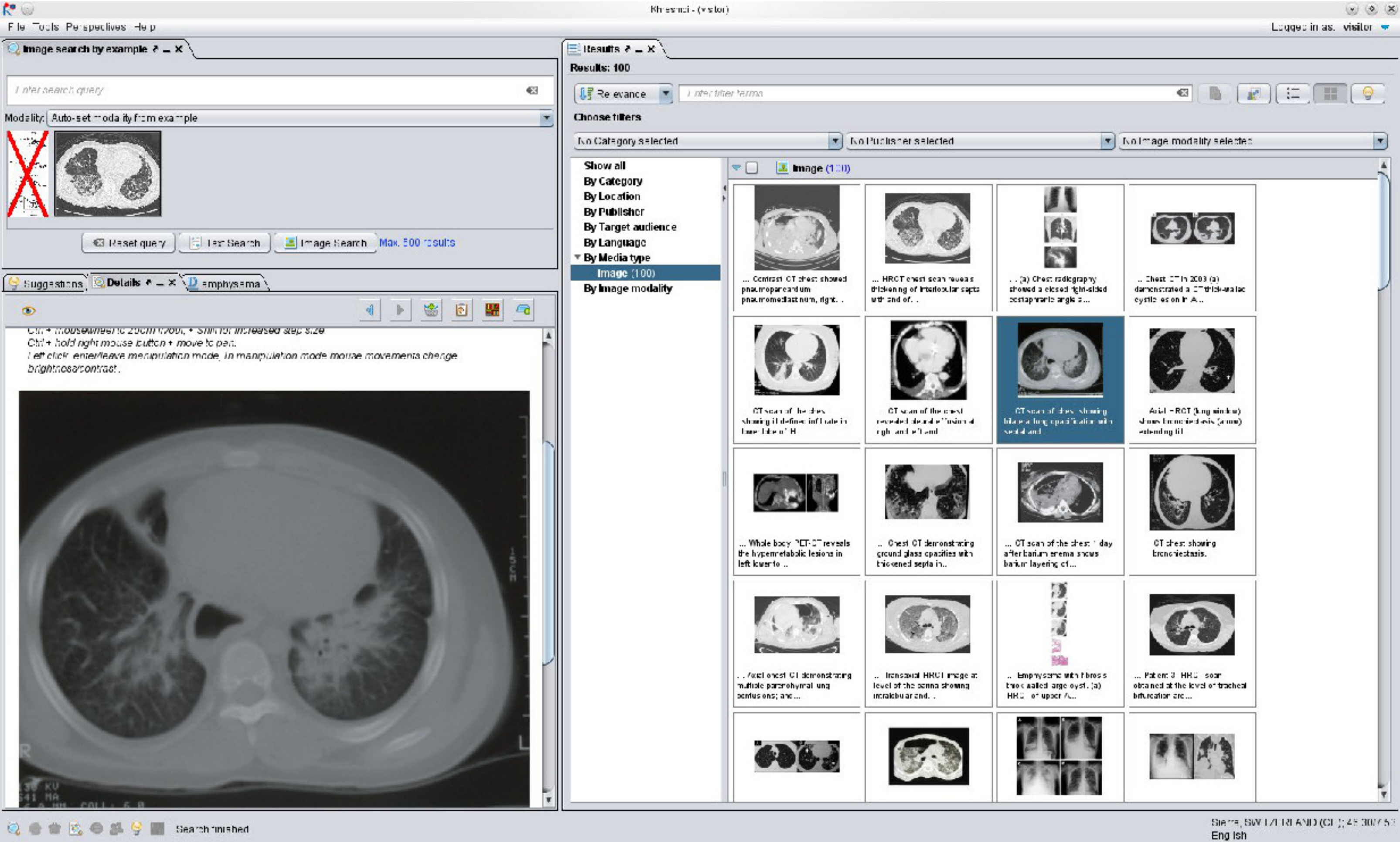}
 \caption{The 2D image retrieval interface.\label{fig:2d_screenshot}}
  \end{center}
 \end{figure}
The user interface of KHRESMOI Radiology is based on ezDL~\cite{BDF2012}.
A more detailed description can be found in~\cite{BDF2014}.
A screenshot of the main 2D image search interface is given in Figure~\ref{fig:2d_screenshot}.
The basic elements are the Query View, the Results View and the Details View.
The user can use the Query view to add text or positive and negative image examples and initiate a search.
Restricting the search with a specific image modality (or a group of modalities) is also supported.

Once a search has been initiated, the results are presented in the Results View in either ranked list or grid format.
Results found in this list can be added in the query to initiate a new search iteration through relevance feedback.
Filtering the results by modalities and media type is also supported.

By selecting a result, its associated information appears in the Details view.
For articles this means the full title, the abstract and the images included.
Search for similar images can be initiated from this view.
For image results this means the full size image, the caption and link to the corresponding article.
Basic image manipulation is available to allow for better image content inspection.

More tools, such as the Personal library and collaborative tools are available and described in more detail in~\cite{BDF2014}.
The indexing and retrieval pipelines that are based on ParaDISE follow below.

\begin{figure}[!tp]
   \begin{center}
     \includegraphics[width=12 cm]{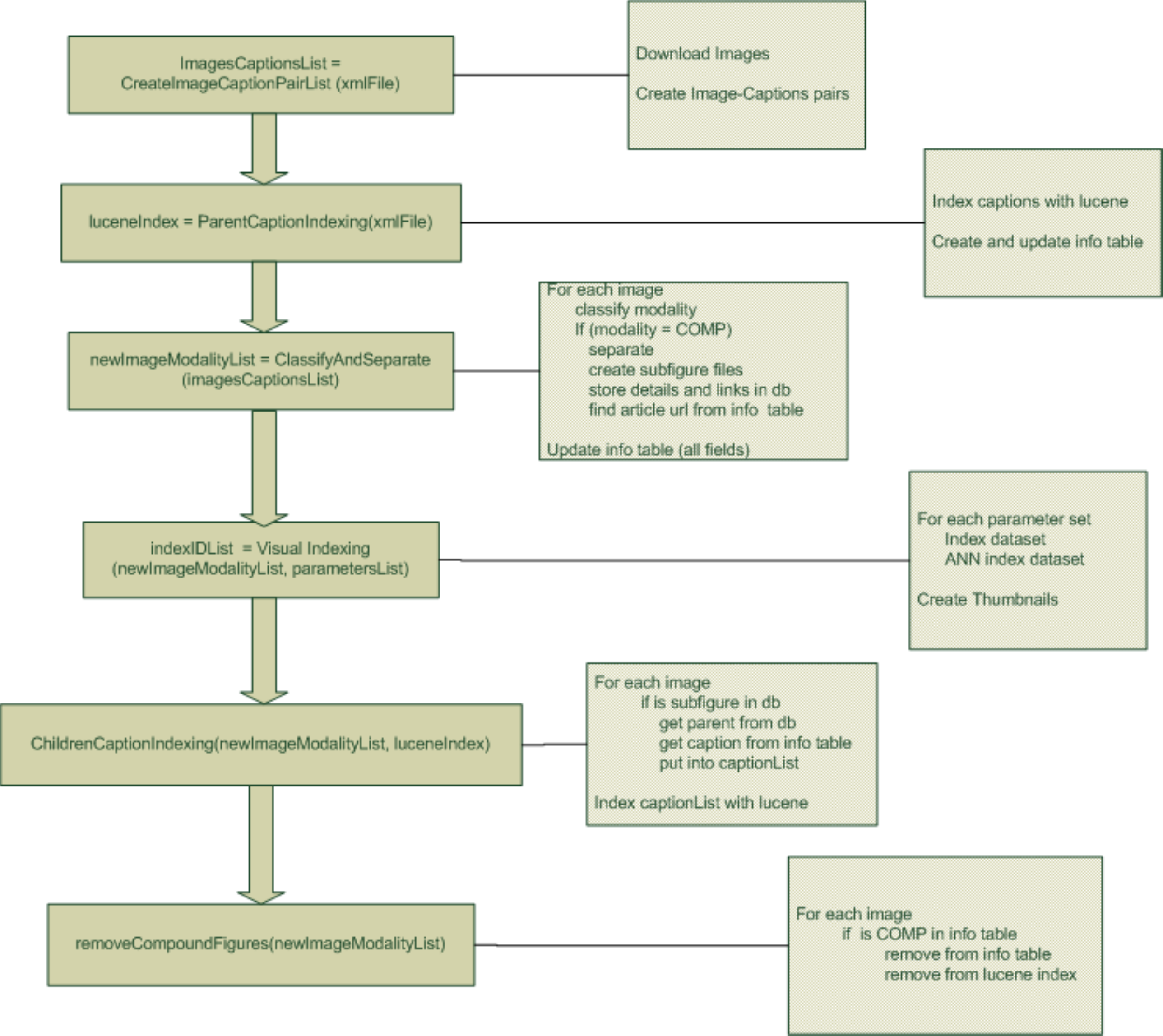}
 \caption{The KHRESMOI indexing pipeline of 2D images.\label{fig:khresmoi_indexing_pipeline}}
  \end{center}
 \end{figure}

In Figure~\ref{fig:khresmoi_indexing_pipeline} the full pipeline of 2D image indexing is presented.
In the beginning, the images are downloaded to the server for faster access and caption--images pairs are created.
Lucene is used to index the captions of the images.
An info table with the various image information, such as the corresponding article URL, the image URL and the caption of the image, is created during that step.

The next step is to classify the images according to their image modality.
The compound figures are separated and their subfigures are saved as new images and are reclassified.
The info table is then updated, including the modality information and the subfigure URLs.
The method presented in~\cite{GMM2013} was used for the modality classification.
The method proposed in~\cite{CMF2013} was used for the compound figure separation.
Hadoop was used for the parallelization of this task.

After a new image list is created with the inclusion of subfigures, ParaDISE undertakes the task of visual indexing.
For the visual indexing, BoVW and Bag--of--Colors (BoC)~\cite{GMM2013} representations were used as shape and color features of the images.
E2LSH was used as an ANN indexing method.

A new round of caption indexing is performed, this time on the subfigures captions.
The compound figures are then removed from the indices.

A dataset of 1.2 million images from 500.000 articles of PubMed Central has been indexed using this pipeline, resulting in 1.7 million images after subfigure indexation.

The development of a new Seeker was dictated by the requirements for the KHRESMOI system, such as search by modality.
The object--oriented implementation of ParaDISE facilitated this and the ModalityFilter Seeker was created.
This Seeker extends the Rocchio Seeker and accepts as input a list of modalities which it uses to filter the results.
The weights used for the Rocchio algorithm are $\beta = 0.6$ and $\gamma = 0.4$.
Query and relevant vectors were considered as the same set of vectors.

\begin{figure}[!tp]
   \begin{center}
     \includegraphics[width=12cm]{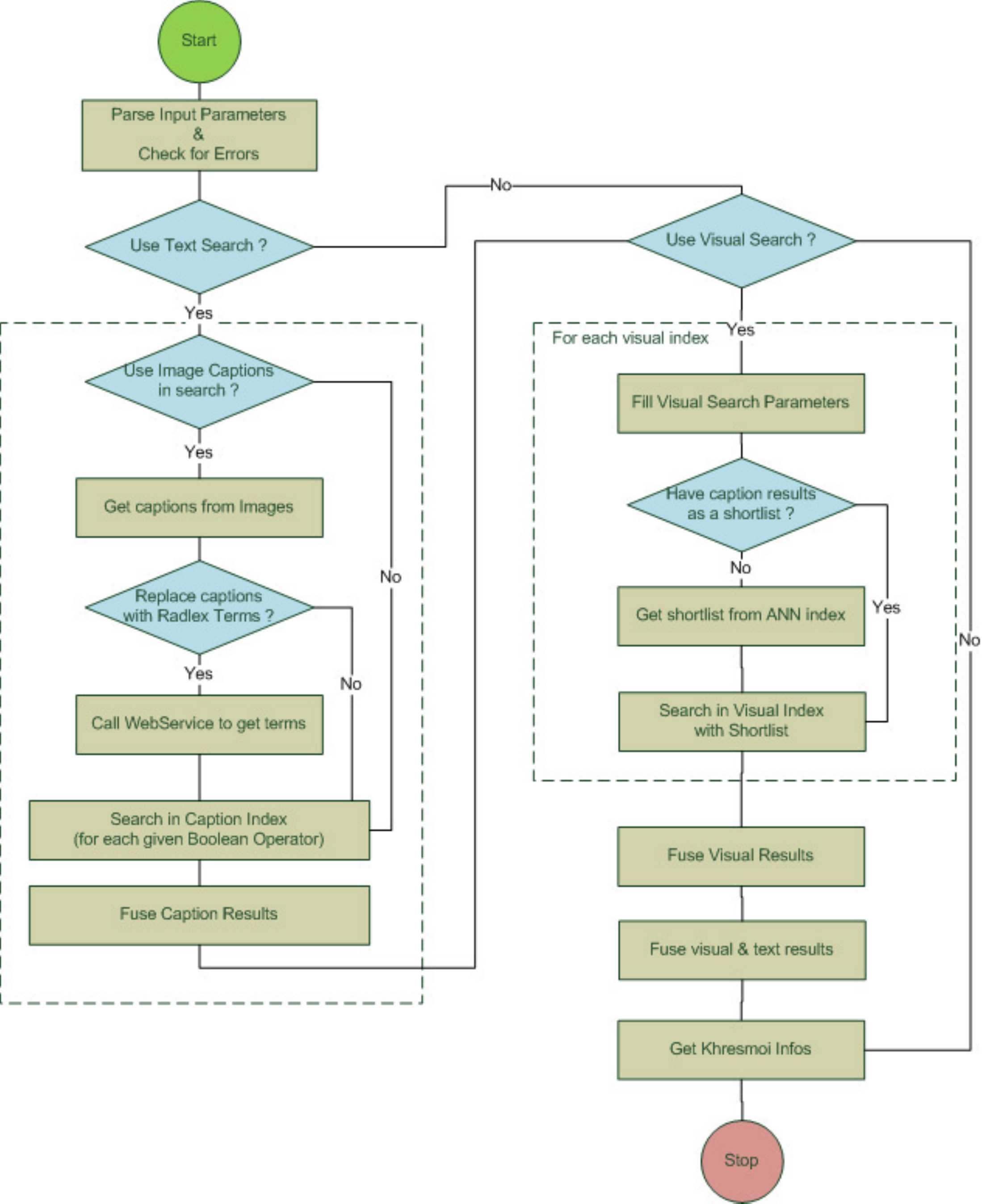}
 \caption{The KHRESMOI search pipeline for 2D image retrieval.\label{fig:retrieval_pipeline}}
  \end{center}
 \end{figure}

The backend 2D image search pipeline is presented in Figure~\ref{fig:retrieval_pipeline}.
Once the Web service is called, the call arguments dictate the behavior of the work flow.
Query Images can be automatically classified to produce a list of target modalities or specific target modalities can be passed as arguments.
If text is included in the query then the text search pipeline is enabled (in the left frame).
Image captions can also be used in relevance feedback iterations.
RadLex terms can be extracted from the captions of the query images using the ONTOtext disambiguation service~\cite{MaT2013} and can be added to the query string.
Captions of negative query image examples have their terms (the ones not present in positive ones) added using the NOT boolean operator.

The next step is the visual similarity search.
For each visual index that needs to be accessed there is a concurrent search using modality filtering.
The histogram intersection similarity measure is used.
If there is no text included in the query, the ANN index is used to build the shortlist to be searched.
Otherwise, the top results returned by the text query constitute the shortlist for the visual search.
The ParaDISE Fusor is then used to fuse the retrieved lists from the visual indices.
The CombMNZ rule is used for this fusion.

The next step consists of the fusion of the text and visual search results, using the Fusor and Reciprocal Rank fusion rule.
Finally, image information existing in the info table is added to the results.

\subsection{Feature Evaluation}

An evaluation on how well visual features (commonly--used in object recognition and scene classification) perform in medical image retrieval was run using ParaDISE.
Two main experiments were run, one for local features and one for global image representations evaluation.
A subset of ImageCLEFmed2012~\footnote{\url{http://www.imageclef.org/2012/medical}}  of 10,000 images was used for this purpose.
First the local features' retrieval performance was evaluated using the BoVW representation for 4 distance measures
(histogram intersection, euclidean distance, cosine similarity and $\chi^2$ distance) and 6 vocabularies of different sizes (10,20,30,40,50, 100).
The BoVW vectors were $l2$ normalized and the Rocchio Seeker was used for the fusion of multiple query images.
The average mean average precision (mAP) over the 6 vocabularies is given in Table~\ref{tab:local_1}.

\begin{table}
\centering
\caption{Average mAP scores of BoVW representations of local features over six vocabularies for different distance measures.}
\label{tab:local_1}
\begin{tabular}{|l|c|c|c|c|}
\hline
Run & HI & L2 norm & Cosine Similarity & $\chi^2$ distance \\
\hline
SIFT & 0.0225  & 0.01087 & 0.0194 & 0.0129 \\
\hline
SURF & 0.0124  & 0.0062 & 0.0081 & 0.0078 \\
\hline
RootSIFT & 0.0207 & 0.0118 & 0.0204 & 0.0140 \\
\hline
Lab & 0.0135 & 0.0127 & 0.0134 & 0.0107 \\
\hline
\end{tabular}
\end{table}

The best performing runs were combined using 4 different fusion rules (CombMNZ, CombSUM, Reciprocal rank fusion and Borda Count) to investigate if they contain complementary information (Table~\ref{tab:local_2}).
The histogram intersection was used for the similarity comparison.

\begin{table}
\centering
\caption{mAP scores of the fusion of best--performing runs: SIFT ($k=20$), SURF ($k=40$), RootSIFT ($k=30$) Lab ($k=100$).}
\label{tab:local_2}
\begin{tabular}{|l|c|}
\hline
fusion Rule & mAP \\
\hline
CombMNZ & 0.0223 \\
\hline
CombSUM  &  0.0216\\
\hline
Reciprocal rank & 0.0206\\
\hline
Borda count & 0.0198 \\
\hline
\end{tabular}
\end{table}

The features were also assessed in 4 different visual vocabulary--based image representations (BoVW, VLAD, SPM and GridBoVW) using the histogram intersection similarity measure (except for the VLAD representation that can have negative values, so cosine similarity was used) (Table~\ref{tab:local_3}).
Small--sized vocabularies were chosen as the dimensionality of VLAD is $k*d$ where $k$ is the number of clusters and $d$ the dimensionality of the feature, so larger vocabularies would result to representations of dimensionality inefficient for quick retrieval.

\begin{table}
\centering
\caption{Average mAP scores of local features for different image representations.}
\label{tab:local_3}
\begin{tabular}{|l|c|c|c|c|}
\hline
Run & BoVW & SPM BoVW & Grid BoVW &  VLAD \\
\hline
SIFT & 0.0225  & 0.0227 & 0.0166 & 0.0181 \\
\hline
SURF & 0.0124  & 0.0123 & 0.0102 & 0.0081 \\
\hline
RootSIFT & 0.0207 & 0.0214 & 0.0158 & 0.0151 \\
\hline
Lab & 0.0135 & 0.0157 & 0.0155 & 0.0050 \\
\hline
\end{tabular}
\end{table}

The best performing local feature is SIFT using all of the distance measures, except cosine similarity where RootSIFT performed slightly better (Table~\ref{tab:local_1}).
It can be seen that the distance metric is very crucial for the retrieval performance.
Similarity measures perform better, with histogram intersection achieving the best results in all the local features.
The fusion of the the best performing runs is not providing better results than the best performing local feature (SIFT) (Table~\ref{tab:local_2}).
This indicates that the evaluated features model the same visual information.

Regarding the local feature representations, SPM appears to enhance the BOVW representation, modelling the spatial information (Table~\ref{tab:local_3}).
Grid spatial modelling degrades performance of BOVW for all features except Lab.
VLAD achieves the worst overall performance, however it is mainly caused by the fact that cosine similarity had to be used instead of histogram intersection.

For the image representations evaluation, 8 descriptors were used.
The two best performing aggregated local feature representations, two global multi--feature descriptors (CEDD, FCTH), two color descriptors (Color layout, Fuzzy color histogram) and two miniature--based descriptors (ColorHoG, GIST).
The results over 4 different distance measures are presented in Table~\ref{tab:global_1}.
The 5 best performing runs (BoVW, SPM BoVW, CEDD, FCTH and Color layout) were combined using CombMNZ to investigate if they contain complementary information.
Histogram intersection was used for this run.

\begin{table}
\centering
\caption{mAP scores of image representations for different distance measures.}
\label{tab:global_1}
\begin{tabular}{|l|c|c|c|c|}
\hline
Run & HI & L2 norm & Cosine Similarity & $\chi^2$ distance \\
\hline
BoVW SIFT k20 & 0.0268 & 0.0107 & 0.0208 & 0.013\\
\hline
SPM BoVW SIFT k40 & 0.0245 & 0.0122 & 0.0109 & 0.0124\\
\hline
CEDD & 0.0216 & 0.010 & 0.020 & 0.0073 \\
\hline
FCTH & 0.0218 & 0.0095 & 0.0207 & 0.009 \\
\hline
Fuzzy Color histogram & 0.0144 & 0.0034 & 0.0152 & 0.0032 \\
\hline
Color Layout & 0.0189 & 0.0134 & 0.018 & 0.0093\\
\hline
ColorHoG & 0.0051 & 0.0063 & 0.005 & 0.0046\\
\hline
GIST & 0.0097 & 0.0014 & 0.0068 & 0.0019\\
\hline
CombMNZ of 5 best & 0.0296 & n/a & n/a & n/a \\
\hline
\end{tabular}
\end{table}

Judging from results of Table~\ref{tab:global_1} the local feature aggregated vectors (BoVW and SPM BoVW using SIFT) achieve the best performance.
Multi--feature descriptors (CEDD and FCTH) come second in performance with Color layout descriptor being the best color histogram.
The miniature--based representations seem to have less consistent mAP even though they perform very well in certain topics.
The distance measure is again shown to be very important in terms of retrieval performance, with histogram intersection and cosine similarity outperforming the Euclidean and $\chi^2$ distance.
The fusion of the best performing runs achieves the highest mAP, indicating this way that the features are complementary.

\section{Discussion}\label{sec:discussion}

The design concepts of ParaDISE were flexibility, expandability and scalability.

Flexibility for such a system is crucial, in order to be usable for both research purposes and as an application.
Evaluating image representations is really important in CBIR as different features perform better for different databases, depending on the content and the task.
Moreover, state--of--the--art CBIR techniques usually include several steps and require a lot of parameter tuning~\cite{YJH2007}.
The choice of component--based architecture for ParaDISE allows for combining local and mid--level features and the evaluation of single steps in the indexing and retrieval pipeline.
The use of editable parameters of the ParaDISE components facilitates tuning parameters and experimenting with different configurations of methods.

Scientific software packages, such as MATLAB can cope with most research tasks.
They are, however, rarely used in practical applications due to their lack of efficiency.
ParaDISE is programmed in JAVA and uses JSON as a data transfer protocol to enable interoperability and realistic application development.
The frontend Web service--based architecture allows for the integration of ParaDISE into larger systems and a flexible hardware topology.
The use of REST and HTTP requests simplifies interaction between the system and various client applications (Web--based or desktop applications that can be written in any language capable of making HTTP requests).

With CBIR being an active research field, novel techniques emerge achieving faster and more precise performance.
Thus, expandability is important to be able to add new components for specific steps or new algorithms for the existing components.
The object--oriented and plugin--like architecture of ParaDISE allows for such expansions (e.g. 3D features, a Classifier component etc.).
The late fusion techniques of the Fusor component can be used to expand the engine by combining it with other retrieval systems (e.g. text--based retrieval engines, such as Lucene).

Last but not least, scalability is a critical issue for many real--life applications and an active research field in CBIR~\cite{AMP2011,PCI2008,JDS2008,SMG2013}.
Indexing large image collections and storing the indices can be troublesome and resource--demanding.
Updating such indices in regular time intervals should be taken into consideration when designing the indexing pipelines.
Moreover, exhaustive search time in large indices is prohibitive in CBIR applications, since CBIR search constitutes of computing distances of image descriptor vectors.
In ParaDISE, parallel indexing is supported using the MapReduce framework~\cite{DeG2008} (see Section~\ref{sec:paradise_index}).
Efficient indexing methods to facilitate fast online search and binary descriptors to reduce memory storage are also supported (Sections~\ref{sec:global_feats_2d},~\ref{sec:paradise_index}, and~\ref{sec:paradise_retrieval}).
The component--based architecture is dealing with scalability by allowing the use of distributed resources and expand when the amount of data and computations grows.

The ParaDISE is available under two different open--source licenses, to facilitate use in commercial applications and research\footnote{\texttt{http://paradise.khresmoi.eu/}}.
The study cases demonstrate the use of ParaDISE in both complex systems but also for evaluation purposes.
The KHRESMOI system has been evaluated by real users, in a user study described in~\cite{MHB2013}.
The results showed high user--satisfaction with aspects such as image and article connection and trustworthiness of results.
Users felt quickly comfortable with CBIR and relevance feedback techniques.

The feature evaluation confirmed the hypothesis that the selection of features is highly dependent to the task.
State--of--the--art local features and image representations such as RootSIFT and VLAD in scene recognition are outperformed by more common descriptors such as SIFT and BoVW.
Moreover, interestingly, global descriptors such as CEDD and FCTH achieve competitive performance.

\section{Conclusions}

ParaDISE is a platform suitable for CBIR or multi--modal retrieval pipelines design, development and evaluation.
Moreover, it can serve as the backend of a standalone application or be integrated into more complex, large--scale systems.
The backend architecture of this system, based on four basic components, make it flexible, distributable and expandable.
The study--cases demonstrate the dual nature of ParaDISE on medical image retrieval, a challenging field in information retrieval,

Future goals include creating an open--source community around ParaDISE that will use and contribute to the development of the platform.
In--house contributions are planned as well, with the inclusion of 3D features and support for region--based retrieval.

\section{Acknowledgements}
This research has received funding from the European Union under grant agreement number 257528 (KHRESMOI).
\bibliographystyle{unsrt}
\bibliography{references}
\end{document}